\documentclass[
    ,final            
  ]
  {aipproc}

\layoutstyle{6x9}

\usepackage{bm}
\usepackage{amssymb,amsmath}

\newcommand{\ket}[1]{|#1 \rangle}
\newcommand{\bra}[1]{\langle #1 |}
\newcommand{\innerproduct}[2]{\langle #1 | #2 \rangle}

\begin{document}

\title{Nuclear dynamics in time-dependent picture}


\author{Takashi~Nakatsukasa}{
  address={Institute of Physics and CCS,
  University of Tsukuba, Tsukuba 305-8571, Japan}
}
\author{Makoto~Ito}{
  address={RI Physics Laboratory, RIKEN, Hirosawa 2-1, Wako 351-0198, Japan}
}
\author{Kazuhiro~Yabana}{
  address={Institute of Physics and CCS,
  University of Tsukuba, Tsukuba 305-8571, Japan}
}

\begin{abstract}
Using the time-dependent theory of quantum mechanics,
we investigate nuclear electric dipole responses.
The time evolution of a wave function is explicitly calculated in
the coordinate-space representation.
The particle continuum is treated with the absorbing boundary condition.
Calculated time-dependent quantities are transformed
into those of familiar energy representation.
We apply the method to a three-body model for $^{11}$Li and to
the mean-field model for $^{22}$O,
then discuss properties of $E1$ response.
\end{abstract}

\maketitle


\section{Time-dependent method for nuclear responses}

Atomic nuclei exhibit a variety of responses to different
experimental probes;
Coulomb and nuclear excitations,
spin- and isospin-dependent probes,
high- and low-energy reactions.
In order to investigate properties of these nuclear excitations,
it is useful to study response function for a probe (external)
operator of interest.
Normally, these quantum many-body problems are theoretically studied
using the energy representation, in which
calculations are carried out by either
diagonalizing the Hamiltonian matrices or
solving differential equations with a fixed energy.
These approaches are advantageous when we are interested
in states within a limited energy range.
However, when we calculate response functions in a wide energy range,
the energy representation may not be the best choice.
Occasionally, the time representation provides more efficient
numerical approaches to quantum problems.
In this paper, instead of taking the
time-independent (energy-dependent) framework,
we present time-dependent approaches to nuclear response functions, 
and show their advantages and usefulness.

\subsection{Time-dependent equation with absorbing boundary}
The quantum mechanical problems are often described by
the eigenvalue equation,
\begin{equation}
\label{E-eq}
\hat{H}\ket{\Psi}=E\ket{\Psi} .
\end{equation}
Since this stationary Schr\"odinger equation is derived from the
time-dependent equation
\begin{equation}
\label{T-eq}
i\hbar \frac{\partial}{\partial t}\ket{\Psi(t)} = \hat{H}\ket{\Psi(t)},
\end{equation}
these two equations should provide the same information if we could
solve Eq. (\ref{E-eq}) for all the energies $E$ and solve Eq. (\ref{T-eq})
for an infinitely long period of time.
However, in practice, we need to truncate either the energy range or
the period of time propagation.
Therefore, these two approaches have advantages and disadvantages,
complimentary to each other.
The stationary equation, Eq. (\ref{E-eq}), is suitable for precise
calculation of eigenstates in a small energy range.
In contrast, the time-dependent one, Eq. (\ref{T-eq}), will be superior
in practice,
when we are interested in a bulk property of response function
for an extended energy range.
The energy resolution obtained in the calculation is, roughly speaking,
inversely proportional to the period of time propagation;
$\Delta E \approx \hbar/T$.
In general, longer the time propagation is, more precise energy
the calculation can predict.
However, for resonance states with positive energies,
the time period $T$ can be safely truncated without much loss
of the accuracy, because the states possess a finite width.

For the positive-energy states,
we must deal with a problem of the boundary condition.
For the time-dependent framework,
since the asymptotic form of the wave function becomes a superposition
of a variety of momentum eigenstates,
solving the time-dependent
Schr\"odinger equation (\ref{T-eq}) with a proper
continuum boundary condition seems to be a very difficult task.
However, the use of the absorbing complex potential provides a
practical approach to the problem \cite{KK86,NB89,SM92,MPNE04}.
We use the complex potential to impose an approximate
outgoing boundary condition on calculated time-dependent wave functions.
This is simply done by replace Eq. (\ref{T-eq}) by
\begin{equation}
\label{T-eq-ABC}
i\hbar \frac{\partial}{\partial t}\ket{\Psi(t)}
= \left(\hat{H}-iW(\vec{r}) \right) \ket{\Psi(t)},
\end{equation}
where $-iW(\vec{r})$ is a coordinate-dependent imaginary potential.
Suppose we calculate the wave function $\ket{\Psi(t)}$ as a function of
time $t$ with an initial
state $\ket{\Psi_\text{in}}$ that is localized in space.
$W(\vec{r})$ must be zero in the region of
$\Psi_\text{in}(\vec{r})\neq 0$.
while it is either zero or positive outside of this region.
In addition, $\nabla W(\vec{r})$ should be small to prevent the reflection,
but simultaneously $W(\vec{r})$ should be large enough to absorb all
the outgoing waves.
As long as these conditions are satisfied, in the region of
$\Psi_\text{in}(\vec{r})\neq 0$,
the wave function, $\Psi(\vec{r},t)$, obtained by solving
Eq. (\ref{T-eq-ABC}) is identical to the solution of Eq. (\ref{T-eq})
with the outgoing boundary condition.
Since small reflection of the waves by the potential $-iW$ is inevitable,
this boundary condition is {\it approximate}, however, we have confirmed
that the approximation is good enough to produce results indistinguishable
from those of the exact boundary condition \cite{NY01,UYN02,NY05}.

\section{Applications}

\subsection{Three-body model: Coulomb breakup of $^{11}$Li}

In this section, we apply the time-dependent method to Coulomb breakup
reaction of the two-neutron-halo nucleus, $^{11}$Li.
In the perturbative regime, properties of the Coulomb excitation
are primarily determined by $dB(E1)/dE$ values.
Strong soft $E1$ excitation in $^{11}$Li were previously reported at
MSU \cite{Ieki93}, at RIKEN \cite{Shim95}, and at GSI \cite{Zins97}.
However, these data seem to be inconsistent with each other.
Recently, the experiment has been performed at RIKEN with much higher
statistics \cite{Nak06}.
The new data indicate a strong $E1$ peak around $E_\text{x}=0.6$ MeV.

The $B(E1)$ strength distribution is defined by
\begin{equation}
\frac{dB(E1)}{dE} = \sum_\mu \sum_{E'} \delta(E-E')
     \left| \bra{\Psi_{E'}} M_{1\mu} \ket{\Psi_0} \right|^2,
\end{equation}
where $\ket{\Psi_0}$ and $\ket{\Psi_E}$ are the ground and excited states
of energy $E$, respectively.
$M_{1\mu}$ is the $E1$ operator with a recoil charge.
This quantity is rewritten in the following way:
\begin{eqnarray}
\frac{dB(E1)}{dE}
&=& \frac{1}{\pi\hbar} \text{Re} \sum_\mu \int_0^\infty dt e^{iEt/\hbar}
          \bra{\Psi_0} M^\dagger_{1\mu} e^{-iHt/\hbar} M_{1\mu} \ket{\Psi_0}
	\\
\label{BE1}
&=& \frac{1}{\pi\hbar} \text{Re} \sum_\mu \int_0^\infty dt e^{iEt/\hbar}
          \innerproduct{\tilde\Psi^\mu(0)}{\tilde\Psi^\mu(t)} .
\end{eqnarray}
Thus, $B(E1)$ strength distribution can be calculated
using the time propagation of the state $\ket{\tilde\Psi^\mu(t)}$.
The initial state is given by
$\ket{\tilde\Psi^\mu(0)}=M_{1\mu} \ket{\Psi_0}$ in Eq. (\ref{BE1}).
It should be noted that, in the time-dependent method,
we do not need to construct the energy eigenstates, $\ket{\Psi_E}$.

We adopt the following Hamiltonian for $^{11}$Li,
neglecting the recoil of the $^9$Li core.
\begin{equation}
H=-\frac{\hbar^2}{2m} \nabla_1^2 -\frac{\hbar^2}{2m} \nabla_2^2
  + V_{nC}(r_1) + V_{nC}(r_2)
  + V_{nn}(|\mathbf{r}_1-\mathbf{r}_2|) ,
\label{H_11Li}
\end{equation}
where $m$ indicates the neutron mass.
The interaction potential between neutron-$^9$Li ($V_{nC}$)
and that of neutron-neutron ($V_{nn}$) are both
in the Woods-Saxon form with radius of 2.3 fm and the diffuseness of 0.6 fm.
Occupied orbitals,
$\phi_s(r_1)$ and $\phi_s(r_2)$, are excluded
from the model space ($H \rightarrow PHP$), using a projection operator,
\begin{equation}
P\equiv \left\{ 1-\ket{\phi_s(1)}\bra{\phi_s(1)}\right\}
\left\{1-\ket{\phi_s(2)}\bra{\phi_s(2)}\right\} .
\end{equation}
The ground state is assumed to be a spin-singlet state, thus
must be symmetric in the coordinate space,
$\Psi_0(\vec{r}_1,\vec{r}_2)=\Psi_0(\vec{r}_2,\vec{r}_1)$.
The depth of $V_{nC}$ potentials are adjusted so as to reproduce
the two-neutron separation energy of 0.3 MeV.
Numerical calculations are performed using the partial-wave expansion
with respect to the $n$-$C$ coordinates, $\vec{r}_1$ and $\vec{r}_2$.
The Lagrange-mesh method is used for the radial coordinate \cite{BH86}.
Since the initial state $\tilde\Psi^\mu(0)$, which decays in time,
is excited in the continuum,
we add the absorbing potentials, $-iW(r_1)$ and $-iW(r_2)$, to
the Hamiltonian of Eq. (\ref{H_11Li}).
These imaginary potentials should be non-zero only where
the initial wave function vanishes ($r_1,r_2>R$).
As the time passes, the two neutrons slowly go away to the infinity
($r_1, r_2 \rightarrow \infty$),
and are absorbed by these potentials.

\begin{figure}[t]
  \label{fig:BE1_11Li}
  \includegraphics[width=0.4\textwidth]{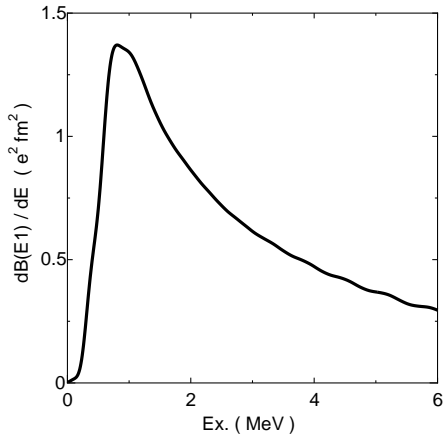}
  \caption{Calculated $B(E1)$ distribution for $^{11}$Li.
  }
\end{figure}

Calculated $E1$ strength distribution,
$dB(E1)/dE$, is shown in Fig.~\ref{fig:BE1_11Li}.
The attractive neutron-neutron interaction, $V_{nn}$,
is chosen to be weaker than
the one to reproduce the $n$-$n$ scattering length in free space
\cite{EBH99}.
The peak position is around $E_\text{x}=0.8$ MeV.
This should be compared to the recent measurement, $E_\text{x}=0.6$ MeV
\cite{Nak06}.
In order to reproduce the observed peak position more precisely,
we may need to improve the Hamiltonian and the ground state of $^{11}$Li.

\subsection{TDHF: Giant dipole resonances in $^{22}$O}

Calculations based on the time-dependent Schr\"odinger equation
of Eq. (\ref{T-eq-ABC}) becomes increasingly difficult
as the particle number increases.
A possible way to avoid this difficulty is to resort to the
density functional (mean-field) theory.
In this section, we use the time-dependent Hartree-Fock (TDHF) method
with the Skyrme interaction to study the dipole response in
neutron-rich Oxygen isotope, $^{22}$O.

Although the TDHF wave packet is capable of describing nuclear
dynamics of the large-amplitude nature,
it is not an easy task to requantize the TDHF trajectories.
Thus, we focus our discussion on its small-amplitude (perturbative) limit.
This is nothing but the well-known random-phase approximation (RPA)
which has been extensively utilized
for studies of both high-energy giant resonances
and low-energy collective excitations \cite{RS80}.
Here, we calculate it in the time-dependent manner.

The RPA equation for the linear response with the density-dependent
interaction is usually derived
from the small-amplitude approximation of
the time-dependent Hartree-Fock (TDHF) theory.
Thus, we should be able to obtain the same information
from the time propagation of a Slater determinant.
Let us assume that the state is initially the ground (HF) state,
then a perturbative external field, $V_\text{ext}(t)$,
is switched on.
Each single-particle orbital follows the TDHF equation
\begin{equation}
i\hbar \frac{\partial}{\partial t}\ket{\psi_i(t)}
= \left(\hat{h}[\rho]+V_\text{ext}(t)\right)\ket{\psi_i(t)} ,
\quad i=1,\cdots,A ,
\end{equation}
where $\hat{h}[\rho]$ is the single-particle Hamiltonian.
The single-particle states are now slightly deviated from its initial
states, $\ket{\phi_i}$:
\begin{equation}
\label{TDHF-state}
\ket{\psi_i(t)}=\left(\ket{\phi_i}+\ket{\delta\psi_i(t)}\right)
e^{-i\epsilon_i t/\hbar} ,
\end{equation}
where $\epsilon_i$ are the HF single-particle energies at the ground state.
Decomposition of $\ket{\delta\psi_i(t)}$ into
normal modes leads to the RPA eigenvalue equations.
Instead of doing this,
we may simply calculate the time-dependent $E1$ dipole moment,
\begin{equation}
\label{T-E1}
D_{1\mu}(t)=\sum_{i=1}^A \bra{\psi_i(t)} M_{1\mu} \ket{\psi_i(t)}
\approx \text{Re} \sum_{i=1}^A \bra{\phi_i} M_{1\mu} \ket{\delta\psi_i(t)} .
\end{equation}
Then, the Fourier transform of Eq. (\ref{T-E1}) provides
information of eigenenergies and transition strength of the RPA normal modes.

We perform the TDHF calculations using the three-dimensional (3D) Cartesian
mesh representation.
The full Skyrme functional with the SGII parameter set is adopted,
which preserves the Galilean invariance.
The time-dependent method provides an efficient tool to calculate the
$E1$ strength function for a wide range of energy.
In fact, to diagonalize the RPA matrix in the 3D coordinate space
is a time-consuming procedure \cite{Muta02,IH03,Ina04}.
The small-amplitude TDHF gives a feasible alternative.

\begin{figure}[t]
  \label{fig:22O}
  \includegraphics[width=.6\textwidth]{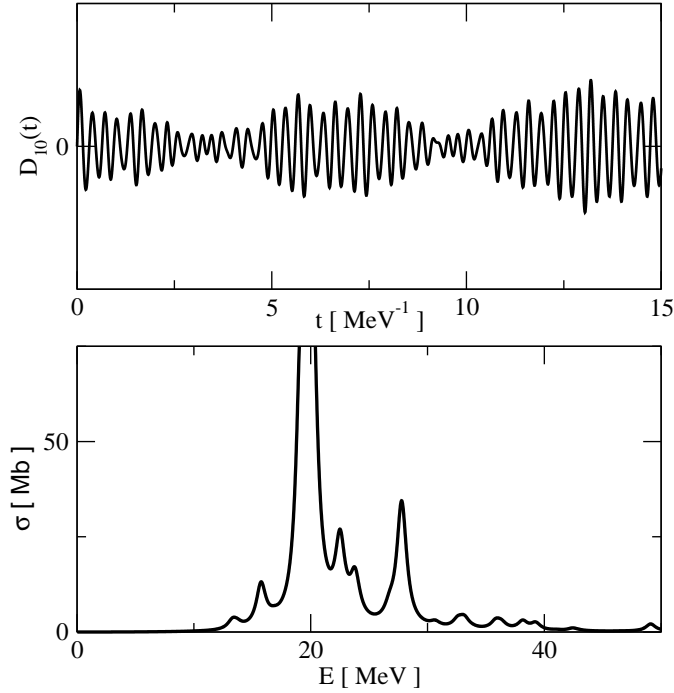}
  \caption{(Top) Calculated time-dependent $E1$ moment for $^{22}$O.
  (Bottom) Calculated photoabsorption cross section for $^{22}$O.
  }
\end{figure}

In the top panel of Fig.~\ref{fig:22O}, we show the time evolution of
$D_{10}(t)$ for $^{22}$O, Eq. (\ref{T-E1}),
with an instantaneous $E1$ external field,
$V_\text{ext}(t)= \eta M_{10}\delta(t)$,
where $\eta$ is an arbitrary parameter.
The time evolution is calculated up to $t=20\ \hbar/$MeV.
Here, we do not use the absorbing potential.
Therefore, the particle continuum is discretized,
leading to discrete peaks in the strength function.
Then, we calculate the Fourier transform of $D_{10}(t)$ with
complex energy, $E+i\Gamma$ with $\Gamma=1$ MeV.
The imaginary part of the energy gives a smoothing width to
every peak.
In the bottom panel of Fig.~\ref{fig:22O},
photoabsorption cross section, calculated in this way, is presented.
The main peak position is almost the same as that of $^{16}$O \cite{NY05}.
Experimental data obtained at GSI suggest a low-energy peak around
$E_\text{x}=9$ MeV \cite{Leis01}.
The present calculation produces the lowest peak at 13 MeV.
This discrepancy may be due to the fact that the neutron separation energy
is too large in the calculation with the SGII parameter set.
The main peak position around $E_\text{x}=20$ MeV
has not been confirmed experimentally, however,
the result is similar to that of the relativistic RPA calculation \cite{VPRL01}.


\section{Summary}

We present the time-dependent method to study nuclear response functions.
The continuum can be properly treated with the absorbing boundary condition.
The method is first applied to a three-body model to investigate
$B(E1)$ distribution in $^{11}$Li.
The calculation produces a strong $E1$ strength at low energy,
however the peak position is slightly too high compared to the experiment.
It seems to be necessary to improve the model Hamiltonian in order to
reproduce the peak position quantitatively.
We next apply the time-dependent Hartree-Fock analysis to
giant resonances.
We show $E1$ resonances in $^{22}$O, using the SGII Skyrme functional.
The main peak is consistent with the relativistic RPA calculation.
The position of the low-energy peak is calculated to be higher
than the experiment by about 4 MeV.
The time-dependent method is an efficient tool to investigate nuclear
response function, especially for its bulk structure in a wide range of energy.




\begin{theacknowledgments}
 This work has been supported by the Grant-in-Aid for Scientific
 Research in Japan (Nos. 17540231 and 17740160).
 The numerical calculations have been performed
 at SIPC, University of Tsukuba,
 at RCNP, Osaka University,
 and at YITP, Kyoto University.
\end{theacknowledgments}



\bibliographystyle{aipproc}   

\bibliography{nuclear_physics,chemical_physics,myself}

\begin{thebibliography}{19}
\expandafter\ifx\csname natexlab\endcsname\relax\def\natexlab#1{#1}\fi
\providecommand{\enquote}[1]{``#1''}
\expandafter\ifx\csname url\endcsname\relax
  \def\url#1{\texttt{#1}}\fi
\expandafter\ifx\csname urlprefix\endcsname\relax\def\urlprefix{URL }\fi
\providecommand{\eprint}[2][]{\url{#2}}

\bibitem[Kosloff and Kosloff(1986)]{KK86}
R.~Kosloff, and D.~Kosloff, \emph{J. Comp. Phys.} \textbf{63}, 363 (1986).

\bibitem[Neuhasuer and Baer(1989)]{NB89}
D.~Neuhasuer, and M.~Baer, \emph{J. Chem. Phys.} \textbf{90}, 4351
  (1989).

\bibitem[Seideman and Miller(1992)]{SM92}
T.~Seideman, and W.~Miller, \emph{J. Chem. Phys.} \textbf{97}, 2499
  (1992).

\bibitem[Muga et~al.(2004)]{MPNE04}
J.~G. Muga, J.~P. Palao, B.~Navarro, and I.~L. Egusquiza, \emph{Phys. Rep.}
  \textbf{395}, 357 (2004).

\bibitem[Nakatsukasa and Yabana(2001)]{NY01}
T.~Nakatsukasa, and K.~Yabana, \emph{J. Chem. Phys.} \textbf{114}, 2550
  (2001).

\bibitem[Ueda et~al.(2002)]{UYN02}
M.~Ueda, K.~Yabana, and T.~Nakatsukasa, \emph{Phys. Rev. C} \textbf{67},
  014606 (2002).

\bibitem[Nakatsukasa and Yabana(2005)]{NY05}
T.~Nakatsukasa, and K.~Yabana, \emph{Phys. Rev. C} \textbf{71}, 024301
(2005).

\bibitem[Ieki et~al.(1993)]{Ieki93}
K.~Ieki, et al., \emph{Phys. Rev. Lett.} \textbf{70}, 730 (1993).

\bibitem[Shimoura et~al.(1995)]{Shim95}
S.~Shimoura, et al., \emph{Phys. Lett. B} \textbf{348}, 29 (1995).

\bibitem[Zinser et~al.(1997)]{Zins97}
M.~Zinser, et al., \emph{Nucl. Phys. A} \textbf{619}, 151 (1997).

\bibitem[Nakamura et~al.(2006)]{Nak06}
T.~Nakamura, et al., \emph{Phys. Rev. Lett.} \textbf{96}, 252502 (2006).

\bibitem[Baye and Heenen(1986)]{BH86}
D.~Baye, and P.-H. Heenen, \emph{J. Phys. A} \textbf{19}, 2041 (1986).

\bibitem[Esbensen et~al.(1997)]{EBH99}
H.~Esbensen, G.~F. Bertsch, and K.~Hencken, \emph{Phys. Rev. C} \textbf{56},
  3054 (1997).

\bibitem[Ring and Schuck(1980)]{RS80}
P.~Ring, and P.~Schuck, \emph{The nuclear many-body problems},
Springer-Verlag, New York, 1980.

\bibitem[Muta et~al.(2002)]{Muta02}
A.~Muta, J.-I. Iwata, Y.~Hashimoto, and K.~Yabana, \emph{Prog. Theor. Phys.}
  \textbf{108}, 1065 (2002).

\bibitem[Imagawa and Hashimoto(2003)]{IH03}
H.~Imagawa, and Y.~Hashimoto, \emph{Phys. Rev. C} \textbf{67}, 037302 (2003).

\bibitem[Inakura et~al.(2004)]{Ina04}
T.~Inakura, et al., \emph{Int. J. Mod. Phys. E} \textbf{13}, 157 (2004).

\bibitem[Leistenschneider et~al.(2001)]{Leis01}
A.~Leistenschneider, et al.,
  \emph{Phys. Rev. Lett.} \textbf{86}, 5442 (2001).

\bibitem[Vretenar et~al.(2001)]{VPRL01}
D.~Vretenar, N.~Paar, P.~Ring, and G.~A. Lalazissis, \emph{Nucl. Phys. A}
  \textbf{692}, 496 (2001).

\end{thebibliography}

\IfFileExists{\jobname.bbl}{}
 {\typeout{}
  \typeout{******************************************}
  \typeout{** Please run "bibtex \jobname" to optain}
  \typeout{** the bibliography and then re-run LaTeX}
  \typeout{** twice to fix the references!}
  \typeout{******************************************}
  \typeout{}
 }

\end{document}